\documentclass[preprint,showpacs,preprintnumbers,amsmath,amssymb]{revtex4}
\usepackage[dvips]{graphicx,color}
\usepackage{dcolumn}
\usepackage{bm}

\def\delsla{\!\not\!\partial}

\begin{document}

\preprint{APT/123-QED}

\title{Lepton effects on the proto-neutron stars 
 with the hadron-quark mixed phase in the Nambu-Jona-Lasinio model}%

\author{Nobutoshi Yasutake$^1$}
 \email{yasutake@th.nao.ac.jp}
\author{Kouji Kashiwa$^2$}
\affiliation{%
$^1$Division of Theoretical Astronomy, National Astronomical Observatory of Japan, 2-21-1 Osawa, Mitaka, Tokyo 181-8588, Japan\\
$^2$Department of Physics, Kyushu University, Fukuoka 810-8560, Japan
}%

\date{\today}

\begin{abstract}
We study the structures of hybrid stars with leptons at finite temperature under beta equilibrium. For the quark phase, we use the three flavor Nambu-Jona-Lasinio~(NJL) model. For the hadron phase, we adopt nuclear equation of state (EOS) by Shen et al.. This EOS is in the framework of the relativistic mean field theory including the tree body effects. For the hadron-quark phase transition, we impose the bulk Gibbs construction or the Maxwell construction to take into account uncertainties by {\it finite size effects}. We find that the pure quark phase does not appear in stable star cores in all cases. With the phase transition, the maximum masses increase $\sim 10 \%$ for high lepton fraction. On the contrary, without the transition, they decrease $\sim 10 \%$. We also find that, in the NJL model, the lepton fraction is more important for structures of unstable stars than the temperature. This result is important for many astrophysical phenomena such as the core collapse of massive stars.
\end{abstract}

\maketitle
\section{\label{sec:level1}Introduction}

The possible existence of quark matter in compact stars has been proposed~\cite{ito,witten}. It should be noted that the uncertainty of EOS is always a big problem in the study of compact star structures. The recent observations imply the existence of large mass compact objects~\cite{ransom05,freire08a,freire08b,verbiest08}. 

The effects of quark matter on various astrophysical phenomena have been studied extensively.
For example, cooling effects on compact star evolutions have been studied in Ref.~\cite{page00, blaschke00, blaschke01, grigorian05}. Other examples include the effects of quark matter on gravitational wave radiation~\cite{lin06, yas07, abdikamalov08}, neutrino emissions~\cite{nakazato08, sagert08}, rotational frequencies~\cite{burgio03}, and the maximum energy release during the transition from neutron stars to quark stars~\cite{yas05, zdunik07}, etc.. However, most of them were based on the MIT bag model.

On the other side, many researchers have proposed EOS's based on finite temperature field theory in beta equilibrium. Among these, the Nambu-Jona-Lasinio (NJL) model is widely  known. The difference between the NJL model and the MIT bag model on the structure of compact stars has been pointed out in Ref.~\cite{menezes06,burgio08}. Furthermore, the NJL model has two notable points. One of them is that the NJL model implies a color-superconducting state, which effects cooling evolution on compact stars with quark matter~\cite{alford08}. The other is that it can show chiral phase transition. We can calculate the spontaneous chiral symmetry breaking by the NJL model. However, baryons in the framework of the NJL model do not reflect experimental result well. This is a big problem in the adaptation of the NJL model for realistic astrophysical phenomena.

To study realistic compact stars including exotic matter,
Burgio and Plumari~(2008) have set two kinds of EOS's for hadron and quark phases~\cite{burgio08}. Hadron EOS in their study was based on Bruekner-Bethe-Goldstone many body theory, which was considered  appropriate for neutron matter~\cite{baldo99}. Furthermore, they adopted the NJL model for the quark phase, and imposed the Maxwell construction on the hadron-quark phase transition. However, there are many uncertainties for the hadron-quark phase transition. Assuming the quark deconfinement transition to be of first order, it causes a thermodynamical instability and the mixed phase appears around the critical density. The properties of the mixed phase depend on electromagnetic and surface contributions to the energy. These contributions are sometimes called "{\it finite-size effects}". The quantitative analysis of these {\it finite size effects} has provided following result:  EOS's for the phase transition become similar to the ones under the bulk Gibbs construction for weak surface tension, and to the ones under the Maxwell construction for strong surface tension~\cite{voskresensky03,endo06,maruyama07}.

In the present paper, we adopt a widely accepted EOS in astrophysics for the hadron phase. The EOS is based on the relativistic mean field~(RMF) theory including many body effects~\cite{shen98}. For the quark phase, we also adopt the NJL model. We impose the bulk Gibbs or the Maxwell construction on the phase transition to take into account uncertainties of {\it finite size effects}. For realistic proto-neutron star~(PNS) structures, the finite temperature effects and neutrino trapping are important. For this reason, in this paper, we also consider the finite temperature case with neutrinos as well.

This paper is organized as follows. In Sec.~II, we outline our EOS's. In Sec.~III, we present numerical result for PNS structures. Sec.~IV is devoted to the conclusion and discussion where we discuss the astrophysical meaning of our result.
\section{Equation of state}
\subsection{\label{sec:level2} Equation of state for quark phase 
---Three flavor Nambu-Jona-Lasinio model}

In a moderately dense system, for example inside compact stars, the quark matter may exist with $s-$quarks. 

The Lagrangian density of the NJL model in the three flavor system is written as   
%
\begin{eqnarray}
 {\cal L} &=& {\bar q}(i \delsla -{\hat m})q + {\cal L}_4 + {\cal L}_6,
\end{eqnarray}
where
%
\begin{eqnarray}
{\cal L}_4 &=&  G_{\rm s} \sum_{k=0}^{8} 
                \Bigl[({\bar q} \lambda^k q)^2 
                     +(i{\bar q} \gamma_5 \lambda^k q)^2\Bigl], \nonumber \\
{\cal L}_6 &=& -K\Bigl[ \det_{i,j}({\bar q}_{i} (1+\gamma_5) q_j) 
                       +\det_{i,j}({\bar q}_{i} (1-\gamma_5) q_j) \Bigr],
\end{eqnarray}
\begin{eqnarray}
q  =  \left(
\begin{array}{c}
q_u\\
q_d\\
q_s\\
\end{array}
\right),~~~
{\hat m} =  \left(
\begin{array}{ccc}
m_u & 0 & 0\\
0 & m_d & 0\\
0 & 0 & m_s\\
\end{array}
\right).
\end{eqnarray}
Here, $i$ and $j$ denote the flavor indices, whereas $q$ and $m_i$ denote quark field and current mass matrix, respectively.
The Lagrangian density components, ${\cal L}_4$ and ${\cal L}_6$ respectively generate four-leg and six-leg interaction.
The six-leg interactions, ${\cal L}_6$, come from $U_A(1)$ anomaly. 
In the chiral limit, $m_f=0$ , the Lagrangian density has 
$SU(3)_L \times SU(3)_R \times SU(3)_{\rm c} \times U(1)_{\rm v}$ 
symmetry.%

With the mean field approximations,
the Lagrangian density is given as
\begin{eqnarray}\hspace*{-5mm}
{\cal L}_{MFA} &=& \sum_{f=i,j,k}    {\bar q}_f(i \delsla - (m_f+\Sigma_{f,{\rm s}}) )q_f - U,
\end{eqnarray}
where
%
\begin{eqnarray}
\Sigma_{i,{\rm s}} &=& -(4G_{\rm s} \langle {\bar q}_i q_i \rangle 
                 -2K \langle {\bar q}_j q_j \rangle \langle {\bar q}_k q_k \rangle ),\\
U &=&  \sum_{l=u,d,s} (2G_{\rm s} \langle {\bar q}_l q_l \rangle ^2 )
          -4K \langle {\bar q}_u q_u \rangle \langle {\bar q}_d q_d \rangle
          \langle {\bar q}_s q_s \rangle,
\end{eqnarray}
for $f\not=j\not=k$.

Therefore, the thermodynamic potential is
%
\begin{eqnarray}
\frac{\Omega}{V} 
      &=&-\frac{T}{V} \ln Z\\
      &=& -2N_c \sum_{f=u,d,s}\int \frac{d^3{\rm p}}{(2\pi)^3}
          \Bigl[E_f ({\rm p}) - \sqrt{{\rm p}^2+m_f^2}  \nonumber \\
      && + \frac{1}{\beta}\ln(1+e^{-\beta E^-_f({\rm p})})
         + \frac{1}{\beta}\ln(1+e^{-\beta E^+_f({\rm p})})
      \Bigr] \nonumber \\
      && + U,
\end{eqnarray}
where $E_f({\rm p})=\sqrt{{\bf p}^2+M_f^2}$, 
$E_f^\pm=E_f({\rm p})\pm \mu_f$ , and $M_f=m_f + \Sigma_{f,{\rm s}}$. 

Since the NJL model is nonrenormalizable, it is necessary to 
introduce a cutoff in the momentum integration. 
In this study, we use the three-dimensional momentum cutoff and
we use the parameter set that is obtained in Ref.~\cite{rehberg96}. 
The cutoff $\Lambda$ is 0.6023 GeV. The coupling constant of four quark interaction is defined as $G_{\rm s} \Lambda^2 = 1.835$, while that of six quark interaction is defined as $K \Lambda^5 = 12.36$.
The current masses of $u-$ and $d-$quarks are fixed at 5.5 MeV and that of $s-$quark is fixed at 140.7 MeV.

The quark number density $n_q$ is given as $n_q=n_u+n_d+n_s= 3 n_B$
with $n_f = \langle q^\dag _f q_f\rangle$ where $n_B$ is the baryon number density.
Moreover, the chiral condensate $\langle {\bar q}_f q_f\rangle$ satisfies the stationary condition 
$\partial \Omega / \partial \langle {\bar q}_f q_f \rangle=0$.

Inside of compact stars, we must impose beta equilibrium and charge neutrality.
Therefore, we rewrite the chemical potential of $u-$, $d-$ and $s-$quarks as 
\begin{eqnarray}
\mu_u &=& \frac{1}{3}\mu_B - \frac{2}{3} \mu_e +\frac{2}{3} \mu_{\nu_e}, \label{muu}\\
\mu_d = \mu_s &=& \frac{1}{3}\mu_B + \frac{1}{3} \mu_e - \frac{1}{3} \mu_{\nu_e},  \label{mud}
\end{eqnarray}
where $\mu_B$ is the effective baryon chemical potential corresponding to the neutron chemical potential. We assume that neutrinos are trapped because it is reasonable for PNS's in the process of core collapse of massive stars.

\begin{figure}
\includegraphics[width=85mm]{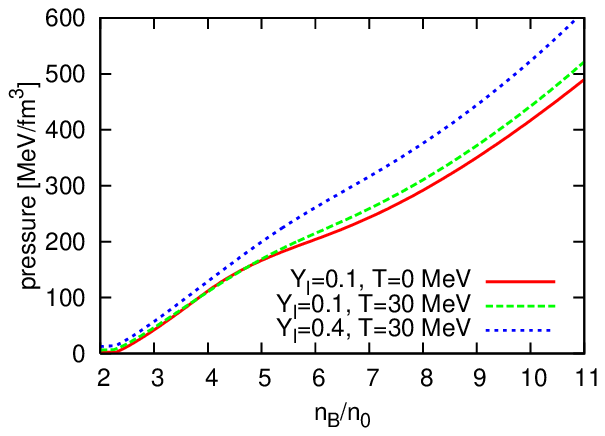}
\includegraphics[width=85mm]{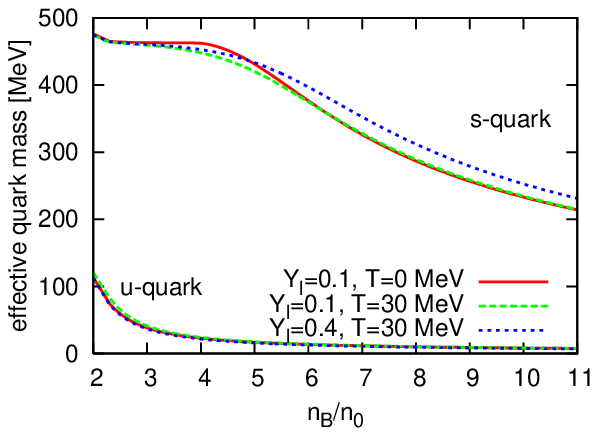}
\caption{\label{fig:eos1} The left panel shows the pressure versus the baryon number density  $n_B$ under beta equilibrium and charge neutrality in the NJL model. The right panel shows the effective quark masses versus $n_B$. Here, we normalized $n_B$ by the nuclear saturation number density $n_0$; $n_0=0.17 fm^{-3}$. We do not show the effective masses of $d-$quark in this figure, because they are almost the same values as the ones of $u-$quark~\cite{menezes06}.}
\end{figure}

Our equation of state in the NJL model is shown in FIG.~\ref{fig:eos1}. 
We set the temperature as $T=0-30$~MeV in this figure, because this is the expected temperature range in compact stars~\cite{burrows86}. For the total lepton fraction, we adopt $Y_l=0.1-0.4$ for the same reason.  The lepton fraction is more important in deciding the stiffness of EOS's than temperature because of the following reason: A high lepton fraction provides a high electron fraction which suppresses the number of other negatively charged particles, such as $s-$quarks, due to the electric charge neutrality condition. This implies that at high lepton fraction case, being the strangeness of content small, the chiral restoration for $s-$quark is suppressed as shown in the right panel of FIG.~\ref{fig:eos1}. Hence, EOS's for high lepton fraction become stiff~\cite{menezes06}. However, the above discussion does not apply to the MIT bag model, because the MIT bag model does not contain the effects of chiral restoration. The details are shown in Menezes et al.(2005)~\cite{menezes06} or Burgio et al.~(2003, 2008)~\cite{burgio03,burgio08}.

An important thing for our EOS is the mass of the $s-$quark and the number density which primarily depend on the chiral restoration, the electric charge neutrality, and the Gibbs/Maxwell construction. However, effects of the color superconductivity may change our result~\cite{bailin84,alford99,iida04,panda04a,panda04b,ruster05,sharma07}. Studies of such effects are out of scope of this work.

In the NJL models, the bag pressures are not free parametric constants as in the MIT bag model.

\subsection{\label{sec:level2} Equation of state for Hadron phase and Mixed phase}
For the EOS of the hadron phase, we adopt the nuclear EOS based on the RMF theory developed by Shen et al.~(Shen EOS)~\cite{shen98}. This EOS includes three-body effects and has been constructed to reproduce the experimental data of masses and radii of stable and unstable nuclei~(see references in Shen et al.(1998)~\cite{shen98}). The range of baryon density is from $10^{5.1}$ g cm$^{-3}$ to $10^{15.4}$ g cm$^{-3}$.

\begin{figure}
\includegraphics[width=85mm]{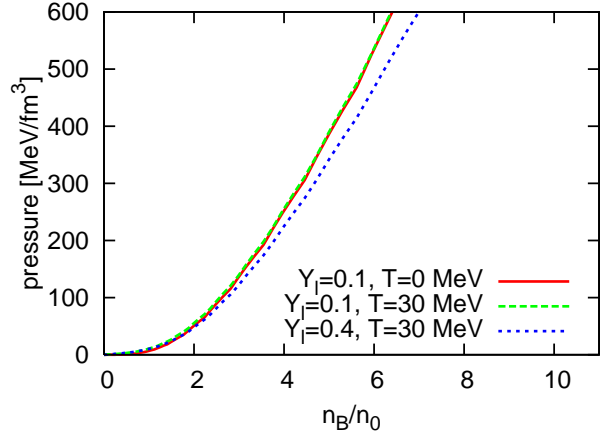}
\caption{\label{fig:02} Same as the left panel of FIG.~\ref{fig:eos1}, but for the nucleon EOS by Shen et al.~\cite{shen98}.}
\end{figure}

Pressure as a function of baryon number density is shown in FIG~\ref{fig:02}. Contrary to the NJL model, this EOS becomes slightly soft for high lepton fraction: e.g. the pressure at $Y_l=0.4$ and $T=30$ MeV is about 10~\% lower than the one at $Y_l=0.1$ and the same temperature. The explanation for this is that, under charge neutrality condition, a high lepton fraction decreases the neutron number density which is the main component of a repulsive nuclear interaction above the saturation density.

For the phase transition, there are many uncertainties which come from {\it finite size effects} as discussed in Sec.~I. In the past studies, it has been shown that EOS's become similar to the ones under the Maxwell construction for strong surface tension, and to the ones under the bulk Gibbs construction for weak surface tension~\cite{voskresensky03,endo06,maruyama07}. Hence, to take into account such uncertainties, we calculate EOS's under the bulk Gibbs construction and the Maxwell construction, respectively. 

First, we describe the bulk Gibbs construction. In this construction, we consider chemical equilibrium at the hadron-quark interface as well as in each phase:
\begin{eqnarray}
\mu_p+\mu_e &=& \mu_n+\mu_{\nu_e}, \label{eq:beta} \\
\mu_n &=& \mu_u+2\mu_d, \label{eq:chemn}\\
\mu_p &=& 2\mu_u+\mu_d, \\
\mu_d &=& \mu_s. \label{eq:ds} 
\end{eqnarray}
Here, $\mu_p$ and $\mu_n$ are the proton chemical potential and the neutron chemical potential, respectively. We assume that $\mu_n$ is the same as $\mu_B$ in equation~(\ref{muu}) and (\ref{mud}). 

The bulk Gibbs constructions impose the following other conditions for phase equilibrium:
\begin{eqnarray}
T^Q  = T^H,~~~P^Q=P^H. \label{eq:TP}
\end{eqnarray}
Here, $T^Q$ and $T^H$ are the temperatures in the quark phase and in the hadron phase, respectively. Similarly, $P^Q$ and $P^H$ denote the pressures in those phases. 

Besides, we consider baryon number conservation law in the mixed phase as,
\begin{eqnarray*}
n_B = \chi n_{B,Q}+(1-\chi) n_{B,H}~~~,
\end{eqnarray*}
where $n_B, n_{B,Q}, n_{B,H}$, and $\chi$ are respectively the total baryon number density, the  baryon number density in hadron matter, the effective baryon number density in quark matter, and the volume fraction which shows quark matter volume divided by the total volume.  Here, $n_{B,Q}$ is one third of the total quark number density. 

Furthermore, we adopt the electric charge neutrality as follows;
\begin{eqnarray*}
Y_e ~ n_B = \chi ~ Y_{p,Q} ~ n_{B,Q}+(1-\chi)~ Y_{p,H}~ n_{B,H}~~~.
\end{eqnarray*}
Here $Y_e$ is the electron fraction, whereas $Y_{p,Q}$ and $ Y_{p,H}$ denote the fraction of positive particles per baryon in quark matter, and in hadron matter. 

Schertler et al.~\cite{schertler99} have calculated the mixed phase under the bulk Gibbs constructions. Their EOS for hadron phase was also based on the RMF theory, and their EOS for quark phase was also the NJL model. However, their calculations were only for zero temperature and did not include neutrinos. 

Under the Maxwell construction, lepton chemical potentials are not considered in the phase equilibrium. The condition of chemical equilibrium for the phase transition is only equation (\ref{eq:chemn}). The Maxwell construction imposes equation (\ref{eq:TP}), too. In the following, we replace $T^H(T^Q)$ and $P^H(P^Q)$ by $T \equiv T^H  = T^Q$ and $P \equiv P^H =P^Q$.

\begin{figure}
\includegraphics[width=85mm]{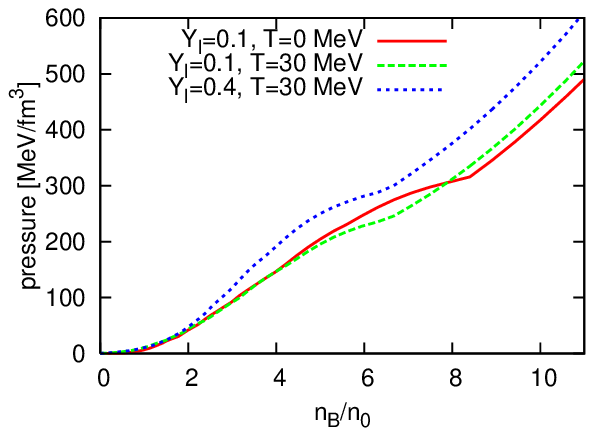}
\includegraphics[width=85mm]{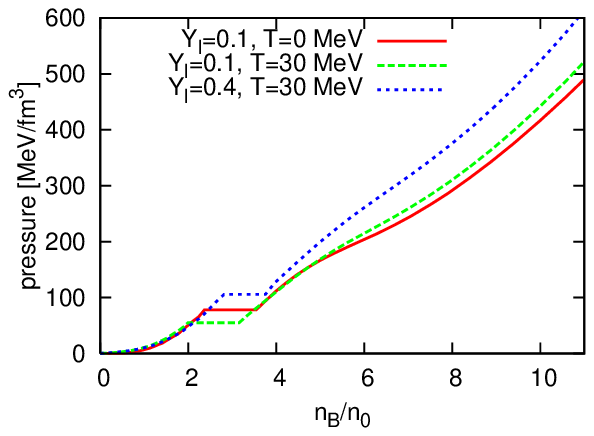}
\caption{\label{fig:eos2} Pressure as a function of the baryon density normalized with the nuclear density under the bulk Gibbs construction~(left panel) and the Maxwell construction~(right panel). }
\end{figure}

Pressure as a function of baryon density is shown in FIG.~\ref{fig:eos2}. In this figure, we set the temperatures and the total electron fractions as $T=0-30$~MeV and $Y_l=0.1-0.4$ as in subsection~A. At these temperatures and lepton fractions, hybrid stars are stable with low central densities, and have maximum masses as we show later in figures of mass-radius relations. Under the Gibbs construction, the density regions of mixed phase are wide; e.g. from 1.19~$n_0$ to 8.40~$n_0$ for $T=0$~MeV and $Y_l=0.1$, where $n_0$ is the nuclear saturation number density given by 0.17 fm$^{-3}$. Such a wide range of mixed phase was also found by Schertler et al~\cite{schertler99}. Under the Maxwell construction, the EOS's have density jumps from the purely hadronic phase to the pure quark phase; e.g. from 2.37~ $n_0$ to 3.54~$n_0$ for $T=0$~MeV and $Y_l=0.1$. We also find that the quark phase appears in low density with increasing temperature, as shown in Burgio et al~\cite{burgio08}. 
%

\section{Structure of hybrid stars}
 We calculate the spherical structures of hybrid stars by Tolman-Oppenheimer-Volkoff (TOV) equation using the EOS's given in Sec.~II. These PNS's have a temperature range of $T=0-30$ MeV, and a lepton fraction range of $Y_l=0.1-0.4$.
 
In FIG.~\ref{fig:mr01}-\ref{fig:mr03}, we display mass versus central density and mass versus radius for compact stars. We assume the dynamical stability condition as 
\begin{eqnarray}
\frac{\partial M}{\partial n_{B,C}} \geq 0,
\label{eq:stable}
\end{eqnarray}
where $M$ and $n_{B,C}$ are the stellar mass and the baryon number density of stellar core, respectively. 

In this paper, we calculate the structures of PNS's at constant temperature for simplicity, not as isentropic matter. The former situation is relevant to the neutrino transparent case~(small $Y_l$) whereas the latter, to the supernova stage (large $Y_l$)~\cite{yasuhira01}.

\subsection{\label{sec:level2}Stable stars}
In this subsection, we discuss the stars satisfying condition~(\ref{eq:stable}) for each EOS. We find that stable stars do not have pure quark phase in their cores for all cases.

\begin{figure*}
\includegraphics[width=180mm]{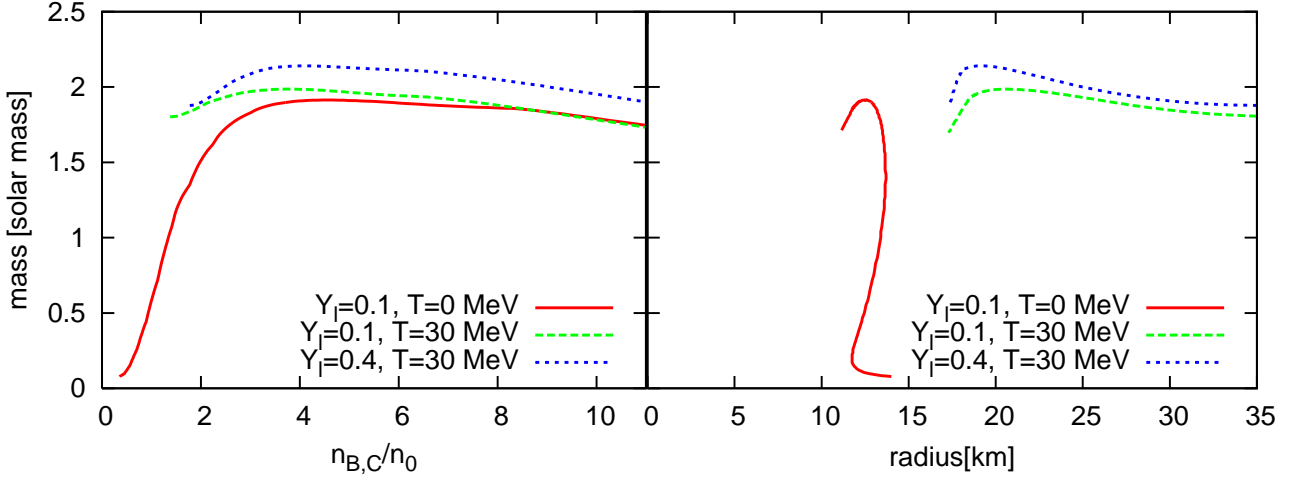}
\caption{\label{fig:mr01} Left panel shows the mass-central density relations of PNS's with the hadron-quark phase transition under the bulk Gibbs construction, and right panel shows their mass-radius relation.}
\end{figure*}

First, we discuss the stable stars with the hadron-quark phase transition under the bulk Gibbs construction. The left panel of FIG.~\ref{fig:mr01} shows the mass-central density relations of PNS's, and the right panel shows their mass-radius relation. Clearly the high lepton fraction~($Y_l=0.4$) enhances the masses because the EOS becomes hard as shown in the left panel of FIG.~\ref{fig:eos2}. The densities at maximum masses are 3.5-5.0 $n_0$, and the matter of cores is in the hadron-quark mixed phase at such densities. The bulk Gibbs construction provides a wide density range for the mixed phase as discussed in Sec.~II. 

\begin{figure*}
\includegraphics[width=180mm]{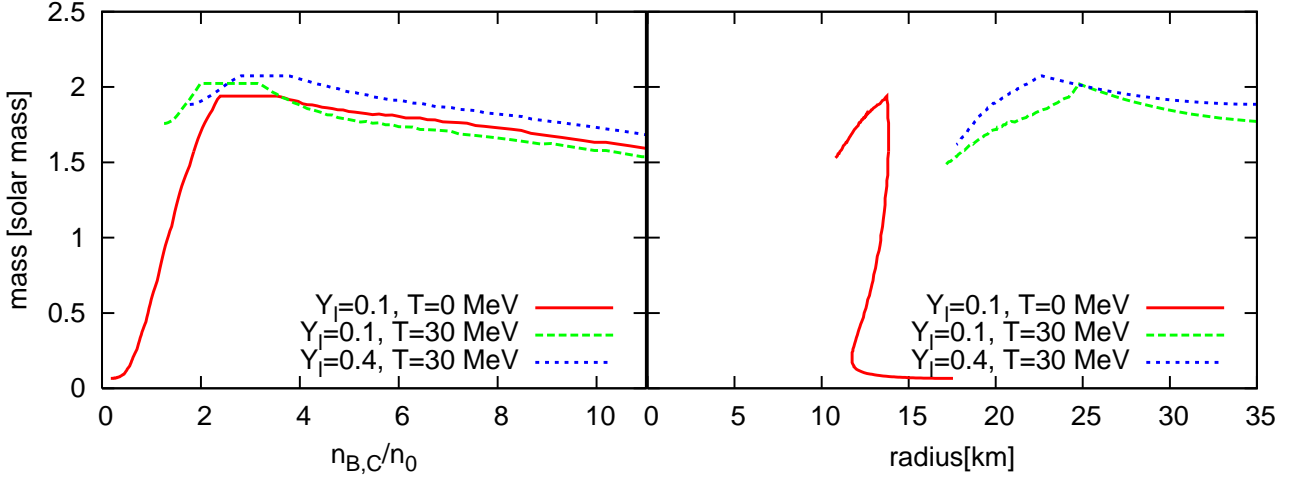}
\caption{\label{fig:mr02} Same as FIG.~\ref{fig:mr01}, but for the Maxwell construction. }
\end{figure*}

Let us move on stable stars with the transition under the Maxwell construction. We find that stable stars do not have pure quark phase in their cores just like the case under the bulk Gibbs construction, as shown in FIG.~\ref{fig:mr02}. At high lepton fraction~($Y_l = 0.4$), the maximum masses become slightly~(3 \%) larger than the ones at low lepton fraction~($Y_l = 0.1$). The explanation for this is that the phase transition occurs under the density where the effects of repulsive nuclear interaction appear on the structures of compact stars. 

There are the central density regions where the gradients of $ \partial M / \partial n_{B,C} $ are zero, since EOS's jump from the hadron phase to the quark phase in these density regions as shown in the right panel of FIG.~\ref{fig:eos2}. 

\begin{figure*}
\includegraphics[width=180mm]{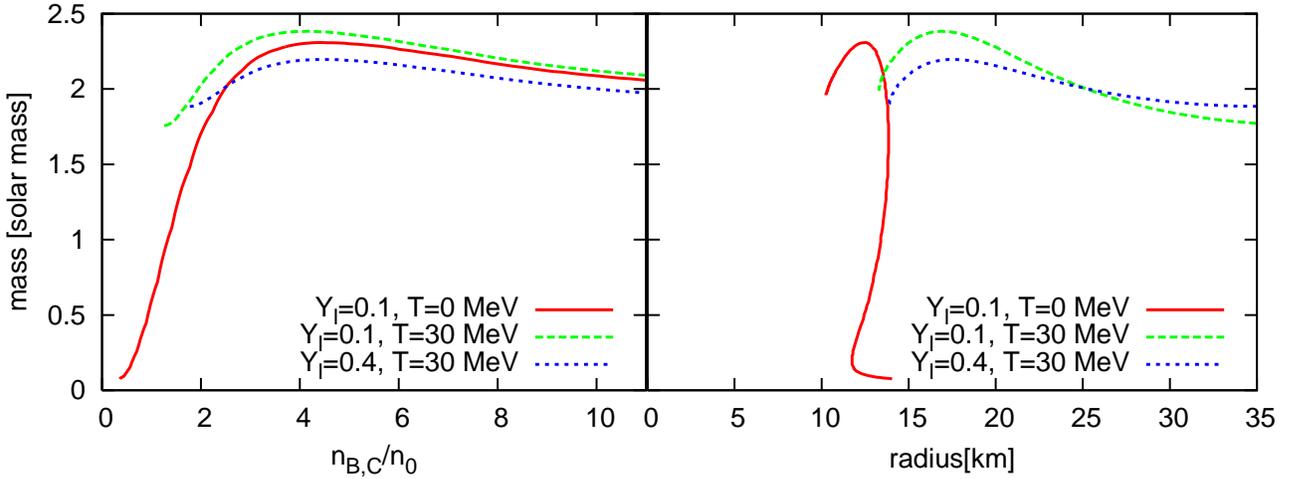}
\caption{\label{fig:mr03} Same as FIG.~\ref{fig:mr01}, but without the hadron-quark phase transition. }
\end{figure*}

On the contrary, without the phase transition, the maximum masses at high $Y_l$ become low clearly~(see FIG.~\ref{fig:mr03}). The explanation is that the EOS's without the transition become soft at high $Y_l$ as discussed in subsection B of Sec.~II.

Maximum masses and the other quantities for each EOS are shown as in TABLE~I.
The label of "bulk Gibbs"~("Maxwell") in the table indicates EOS with the hadron-quark phase transition under the bulk Gibbs~(the Maxwell)  construction, whereas the label of "Shen" indicates without the transition, adopting only Shen EOS.
 
For the EOS's with the transition, the maximum masses become large for high lepton fractions; e.g. for the bulk Gibbs construction, the maximum masses for $Y_l=0.1$ are in a range around 1.95 solar masses. Those for $Y_l=0.4$ are in another range around 2.10 solar masses which are 8 \% larger than the ones for $Y_l=0.1$. For the Maxwell construction, the maximum masses for $Y_l = 0.4$ are 3 \% larger than the ones for $Y_l=0.1$. In this case, the typical densities of compact stars are $1-2 n_0$ because the maximum densities for stable stars are $\sim 2 n_0$. For such densities, the effects of repulsive nuclear interaction at low lepton fraction do not appear~(see FIG.~\ref{fig:02}).

On the contrary, the maximum masses without the phase transition for $Y_l =0.4$ are 9\%  smaller than the ones for $Y_l =0.1$. This is exactly the opposite of the result obtained with the phase transition. The following observations help us to understand this difference; the maximum densities for stable stars are $\sim 4.5 n_0$. For such densities, the EOS's without the transition become soft at high $Y_l$ as discussed in subsection B of Sec.~II~(see FIG.~\ref{fig:02}).

Let us discuss the evolution of compact stars. We calculate structures of hydro static compact stars at a wide range of temperature and lepton fractions. Here, we assume that these sets of stars are snapshots of PNS evolutions. Let us change the $T$ and $Y_l$  values from $T=30$~MeV, $Y_l=0.4$ to $T=0$~MeV, $Y_l=0.1$. The maximum mass difference $\delta M_{max}$ between these two cases is $\sim 0.23 M_\odot$~($\sim 0.13 M_\odot$) with the transition under the bulk Gibbs~(the Maxwell) construction. On the contrary, the Shen EOS show opposite behavior  in maximum masses (i.e. $\delta M_{max} \sim -0.11 M_\odot$). This difference implies that the EOS's with the phase transition become soft after deleptonization although the EOS's become hard  without the transition. The observations of neutrinos for core collapse supernovae will  show us which EOS is acceptable in the future. 

\begin{table}
\caption{\label{tab:maxmass}
Maximum masses in FIG.~\ref{fig:mr01}-\ref{fig:mr03}. The label of "bulk Gibbs"~("Maxwell") indicates EOS with the hadron-quark phase transition under the bulk Gibbs~(the Maxwell)  construction, whereas the label of "Shen" indicates without the transition~(adopting only Shen EOS).
}
\begin{ruledtabular}
\begin{tabular}{lcccc}
$Y_l$ & $T$~(MeV) & $M_{max}$/$M_\odot$ & $R$~(km) & $n_{B,c}/n_0$ \\
\hline
\multicolumn{5}{c}{bulk Gibbs} \\
\hline                                  
 0.1 &  0 & 1.91 & 12.5 & 4.56 \\  
     & 30 & 1.99 & 20.7 & 3.71 \\     
\hline
 0.4 &  0  & 2.05 & 12.7 & 4.78 \\  
      & 30 & 2.14 & 19.2 & 4.07 \\    
\hline
\multicolumn{5}{c}{Maxwell} \\
\hline                                   
 0.1 &  0 & 1.94 & 13.8 &  2.39\\     
     & 30 & 2.02 & 24.7 &  2.03\\   
\hline
 0.4 &  0  & 1.99 & 13.4 & 3.38\\ 
      & 30 & 2.07 & 22.7 & 2.80\\                                     
\hline
\multicolumn{5}{c}{Shen} \\
\hline
0.1 &  0 & 2.31 & 12.5 &  4.44\\     
     & 30 & 2.38 & 16.9 &  4.16\\   
\hline
 0.4 &  0  & 2.11 & 12.3 & 5.00\\ 
      & 30 & 2.20 & 17.6 & 4.46\\           
\end{tabular}
\end{ruledtabular}
\end{table}
%
\subsection{\label{sec:level2}Unstable stars}
In this subsection, we consider unstable stars which do not satisfy the condition~(\ref{eq:stable}).

At $n_{B,C}/n_0 \sim 10$, all stars are unstable. Such stars collapse to black holes. It is clear that a lepton fraction is more important in deciding a stellar mass than temperature~(see FIG.~\ref{fig:mr01}-\ref{fig:mr03}). The masses with the transition under the bulk Gibbs~(the Maxwell) construction are about 1.8~$M_\odot$~(1.6~$M_\odot$) for $Y_l=0.1$ at any temperature and  $n_{B,C}/n_0=10$, whereas they are about 2.0~$M_\odot$~(1.7~$M_\odot$) for $Y_l=0.4$. This is because the pure quark phase appears in these stellar cores. The stiffness of EOS in the NJL model is decided by lepton fraction, but not by temperature as described in Sec.~II~(see FIG.~\ref{fig:eos1}). On the contrary, the masses without the transition are about 2.1~$M_\odot$ for $Y_l=0.1$, and 2.0~$M_\odot$ for $Y_l=0.4$. 

Therefore, these lepton effects on EOS's will be important for some astrophysical phenomena, such as black hole formations. Our prediction is that trapping of neutrinos will suppress the speed of the collapse because  the EOS in the NJL model is more stiff. However, temperature will not provide such suppression.

We note that our calculation is only for spherical stars. It is not clear whether the stars are really stable or not when rotation and magnetic field are also included~\cite{kiuchi08}.

\section{Conclusion and Discussion}
We study the structures of PNS's with leptons at finite temperature considering the hadron-quark phase transition. To take into account {\it finite size effects}, we impose the Gibbs or the Maxwell constructions on the phase transition. Our studies show that the pure quark phase does not appear in stable stellar cores for all cases.

The maximum masses with high lepton fraction ($Y_l=0.4$) become 8~\% higher than those with low electron fraction ($Y_l=0.1$) with hadron-quark phase transition under the bulk Gibbs construction, because the effects on the NJL model, which stiffen EOS's for high lepton fraction, appear in the mixed phase.

Under the Maxwell construction, the maximum masses with high lepton fraction ($Y_l=0.4$) become 3~\% higher than those with low electron fraction ($Y_l=0.1$), because the phase transition occurs under the density where the effects of repulsive nuclear interaction appear at low lepton fraction.

On the contrary, without the transition, the maximum masses with high lepton fraction become 9~\% lower than the ones with low electron fraction, because the hadron EOS which we adopt is soft at high lepton fraction.

Hence, we conclude that the EOS's with the phase transition become soft after deleptonization, although the EOS without the transition becomes hard. 

We also find that lepton fraction is more important to determine structures than temperature for unstable stars. The explanation for this is that the pure quark phase in the NJL model appears in these stelar cores, where the EOS's become hard for high lepton fraction, because the chiral restoration for $s-$quark is suppressed. On the contrary, the EOS's without the transition become soft for high lepton fraction as discussed in the last paragraph. These behaviors are important for some astrophysical phenomena such as core collapse supernovae, since the effect of leptons on EOS's changes many dynamical aspects~\cite{nakazato08,kotake03,buras06,janka07,ott08}.

We note that EOS's have many uncertainties. For the hadron phase, we do not take into account hyperons in this paper. The effects of hyperons on the phase transition are out of the scope of this work and will be presented elsewhere. For quark matter, there are uncertainties in the NJL model itself, such as higher order interaction effects~\cite{kashiwa07,kashiwa07b} and strong magnetic field~\cite{fukushima08,jorge07}. Moreover, there are other extended NJL models, such as PNJL model~\cite{fukushima08a, fukushima08b}. These are open questions for astrophysics and nuclear physics.

We only use the TOV equation to solve stellar structures. This equation describes stellar structures at hydrostatic equilibrium, however it is not clear whether the stars are thermodynamically stable or not under the same condition. 
In other words, it does not include any effects of thermodynamical convections. Dynamical simulations or linear analysises will provide further insight of the effects of those convections.

\begin{acknowledgments}
We would like to thank Toshitaka Tatsumi, Toshiki Maruyama, Hiroki Kouno, Masayuki Matsuzaki, and Masanobu Yahiro for useful discussions and suggestions. We are also grateful to Yolande McLean and Yamac Pehlivan for reading the manuscript.
\end{acknowledgments}

\newpage 
\bibliography{yasE}

\end{document}